# MutaGAN: A Seq2seq GAN Framework to Predict Mutations of Evolving Protein Populations

Daniel S. Berman*, Craig Howser*, Thomas Mehoke, Jared D. Evans

*These authors contributed equally to this work.

E-mail address of corresponding author: Daniel.Berman@jhuapl.edu

Johns Hopkins Applied Physics Laboratory, 11100 Johns Hopkins Rd., Laurel, MD 20723

## Abstract

The ability to predict the evolution of a pathogen would significantly improve the ability to control, prevent, and treat disease. Despite significant progress in other problem spaces, deep learning has yet to contribute to the issue of predicting mutations of evolving populations. To address this gap, we developed a novel machine learning framework using generative adversarial networks (GANs) with recurrent neural networks (RNNs) to accurately predict genetic mutations and evolution of future biological populations. Using a generalized time-reversible phylogenetic model of protein evolution with bootstrapped maximum likelihood tree estimation, we trained a sequence-to-sequence generator within an adversarial framework, named MutaGAN, to generate complete protein sequences augmented with possible mutations of future virus populations. Influenza virus sequences were identified as an ideal test case for this deep learning framework because it is a significant human pathogen with new strains emerging annually and global surveillance efforts have generated a large amount of publicly available data from the National Center for Biotechnology Information's (NCBI) Influenza Virus Resource (IVR). MutaGAN generated "child" sequences from a given "parent" protein sequence with a median Levenshtein distance of 2.00 amino acids. Additionally, the generator was able to augment the majority of parent proteins with at least one mutation identified within the global influenza virus population. These results demonstrate the power of the MutaGAN framework to aid in pathogen forecasting with implications for broad utility in evolutionary prediction for any protein population.

Keywords: Generative Adversarial Networks; Sequence Generation; Influenza Virus; Deep Learning; Evolution

## 1 Introduction

Biological evolution manifests itself through random mutations that occur during genome replication. These mutations cause changes in protein sequence and structure which can affect the ability of an organism to interact with its environment. When this change improves organismal fitness, the probability the mutation is passed on to future generations is increased. Virus replication is inherently error-prone, and only mutations that maintain the ability to infect hosts and evade the host immune system are inherited by subsequent generations. Mutations that reduce fitness do not persist and these variants are ultimately lost from the population. Because these mutations occur randomly in the genetic sequence that codes for these proteins, and because it is difficult to predict which mutations will lead to improved fitness, it is difficult to predict which strains will emerge and become predominant.

Although it is not currently possible to capture all variables that give rise to population emergent traits, modeling the appearance and stability of different mutations over time can serve as a proxy for understanding environmental pressures [1] [2] [3] [4]. Subsequently, if an accurate model can be created, changes that occur in future populations can be predicted [5] [6] [7]. Tools to predict the evolution of a biological organism would significantly improve our ability to prevent and treat disease. The knowledge of how an organism will evolve would allow us to develop more precise interventions and preventive measures in advance and to prevent outbreaks or combat invasive species. Deep learning has led to performance breakthroughs in a number of applications but has yet to contribute toward predicting mutations and evolution of biological populations. We viewed this problem of predicting mutations as analogous to some natural language processing (NLP) tasks, like translation and text generation, for which deep learning has proven successful, making it a great model candidate.

In this paper, we propose MutaGAN, a novel deep learning framework that utilizes generative adversarial networks (GANs) and sequence-to-sequence (seq2seq) models to learn a generalized time-reversible evolutionary model. We then demonstrate its capability of accurately modeling the mutations observed in phylogenetic data of the H3N2 influenza virus hemagglutinin (HA) protein. To our knowledge, this is the first deep learning model that attempts to model and predict the evolution of a protein with minimal human input and no human supervision.

## 2 Background

The focus of this paper is on the machine learning aspects of MutaGAN. Therefore, we briefly review the application of deep learning to biological sequences and the deep learning algorithms and techniques used in this paper. A more detailed and nuanced review of biological sequences and amino acids is out of the scope of this paper.

### 2.1 Deep Learning and Biological Sequences

New methods in data science have been applied to biological sequences for purposes of unsupervised characterization and supervised classification tasks. Deep learning is a natural candidate for these efforts due to an exceptional ability to abstract higher order structure from high-resolution and complex datasets. Previous work has applied natural language processing (NLP) techniques to genomic sequence sets [8] [9] [10] [11] [12] [13]. Ng created a word embeddings process for DNA, called dna2vec, which creates vector representations for short substrings of DNA sequences [14]. The extension of deep neural network architectures such as convolutional neural networks (CNNs), recurrent neural networks (RNNs), and stacked autoencoders onto biological sequence data has proven useful for DNA sequence classification [15] [16] [17] as well as prediction of RNA binding-sites [18], protein-protein interactions [19], and DNA-protein binding [20]. Furthermore, deep learning methods have been extended to the problem of protein-folding [21] [22] to predict molecular characteristics like secondary structure [23], backbone angle and solvent accessibility surface areas [24], and other details about proteins [25].

In 2014, Goodfellow et al. developed a technique for training generative models called GANs [26] [27]. GANs have seen the greatest success in image generation [28] [29] [30] [31] and have also been used to generate text [32] [33] [34] [35]. GANs have also been extended to bioengineering applications, where they were implemented in conjunction with CNNs to optimize DNA for microarray probe design [36], protein sequences for discovery of novel enzymes [37], as well as implemented with an RNN for gene sequence optimization for antimicrobial peptide production [38], all from random noise. Additionally, a CNN-based GAN was used to predict the most probable folding of protein sequences given amino acid sequence and pairwise distances between α-carbons on the protein backbone [39]. In all of these cases, the sequences corresponded to between 50 and 300 amino acids. However, none of these used an RNN conditional GAN to model the natural evolution of a biological sequence.

## 2.2 Sequence-To-Sequence Model

The specific deep learning architecture used to enable high performance encoded representations of amino acid sequences is known as a sequence-to-sequence model (seq2seq). A seq2seq model is a type of neural machine translation algorithm that uses at least two recurrent neural networks (RNNs), like long-short term memory (LSTMs) [40], that take as input a sequence with the goal of constructing a new sequence [41]. There are two parts to this model: an encoder and a decoder, shown in Figure 1. The encoder $E$ takes as input a sequence and converts it into a vector of real numbers. This vector is then used as the initial state of the decoder, which constructs the goal sequence. Seq2seq models have shown success in translation tasks [42] [43] and text summarization tasks [44]. For this reason, we viewed the problem of modeling protein evolution from parent to child as a translation problem. The seq2seq model in MutaGAN uses a bidirectional encoder [45], simultaneously evaluating the input sequence forwards and backwards to produce the optimally encoded vector.

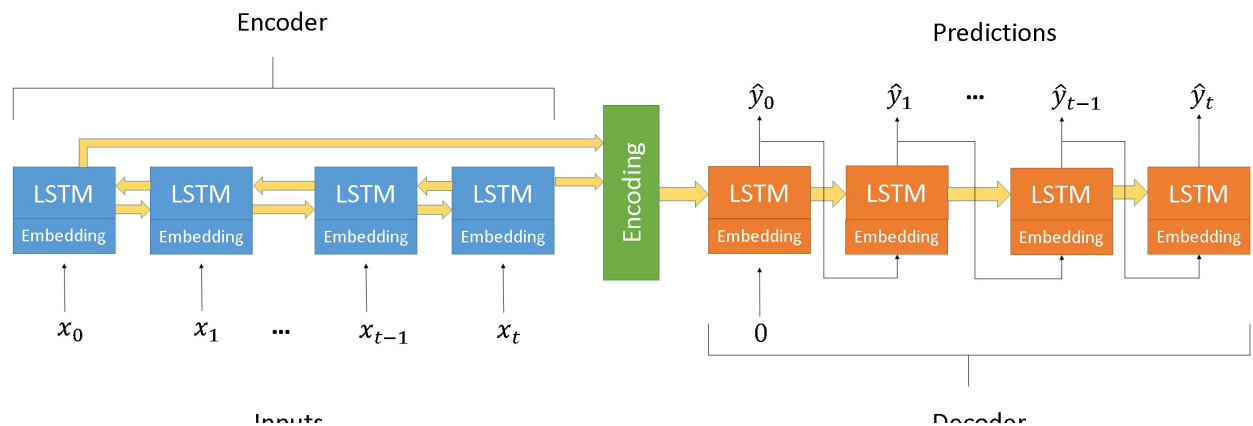

**Figure 1: A seq2seq model using two bidirectional LSTMs encoder and a unidirectional LSTM decoder and embedding layers.**

## 2.3 GANs

A GAN consists of two neural networks, a generator *G* and a discriminator *D*, that compete in a zero-sum game of the generator trying to fool the discriminator and the discriminator trying to distinguish real examples from generated examples. The traditional methodology for training a GAN alternates between training the discriminator and freezing the weights in the discriminator and training the GAN to generate sequences that the discriminator thinks are real. Typically, a GAN is trained to turn random noise into an output matching a known distribution. However, by conditioning the output of a GAN on a partially structured input, in addition to random noise, we can implement a conditional GAN [27]. The conditional GAN used in this paper is shown in Figure 2. In the context of this work, our partially structured input was the parent protein sequence which, once encoded, was combined with a random vector of noise. Because mutations are inherently stochastic, we identified a conditional GAN framework as the ideal model candidate for the use of a seq2seq model to generate numerous mutations given a single parent protein.

## 3 MutaGAN

The core of our model was a seq2seq translation deep neural network, which formed the generator in the GAN. The seq2seq encoder *E* takes as input a sequence of length *N*, with a dictionary size *d*, and converts it into a vector of length *m*, $E: \mathbb{R}^{N \times d} \to \mathbb{R}^{1 \times m}$, analogous to the embedding layer. The embedding layer of this network was created using a biological language of 3-mers of amino acids with a sliding window with step size of 1. For example, the sequence of amino acids *MKTIIALSY* is transformed into *MKT KTI TII IIA* …. The output of the decoder LSTM is fed into a softmax dense layer (Figure 2). To achieve our goal of a model that can generate different sequences for a given parent, the random noise vector was combined with the output of the encoder.

The structure of the encoder in the discriminator is slightly different from the encoder in the generator, but it uses the same weights. An embedding layer requires the input to be in the form of a single integer, representing a discrete input. However, the output of the generator at each time step is a probability vector with dimension $\mathbb{R}^{d \times 1}$. This cannot be transformed into an integer with the argmax function because the argmax function is not differentiable, meaning it does not allow for backpropagation to train the generator. Therefore, a modified encoder takes as input a vector with dimensions $\mathbb{R}^{d \times 1}$, with the first layer of the modified encoder being a linear dense layer with an output of *m* and no bias term. The weights of this dense layer are the same as those of the embedding layer, meaning it produces a linear combination of the embeddings from the embedding layer of the encoder for parent sequences in the discriminator and the encoder in the generator. The generated sequences were fed into this encoder as softmax outputs of the generator and the real sequences were fed in as one-hot encoded sequences.

The architecture of the discriminator was built using code shown in Supplemental Table 1, with the final layer of the discriminator being a sigmoid function. This includes the loading of the pretrained autoencoder weights. Because the two encoders used the same bidirectional LSTM, the weights for that in the two encoders were automatically shared once they were loaded into the parent encoder.

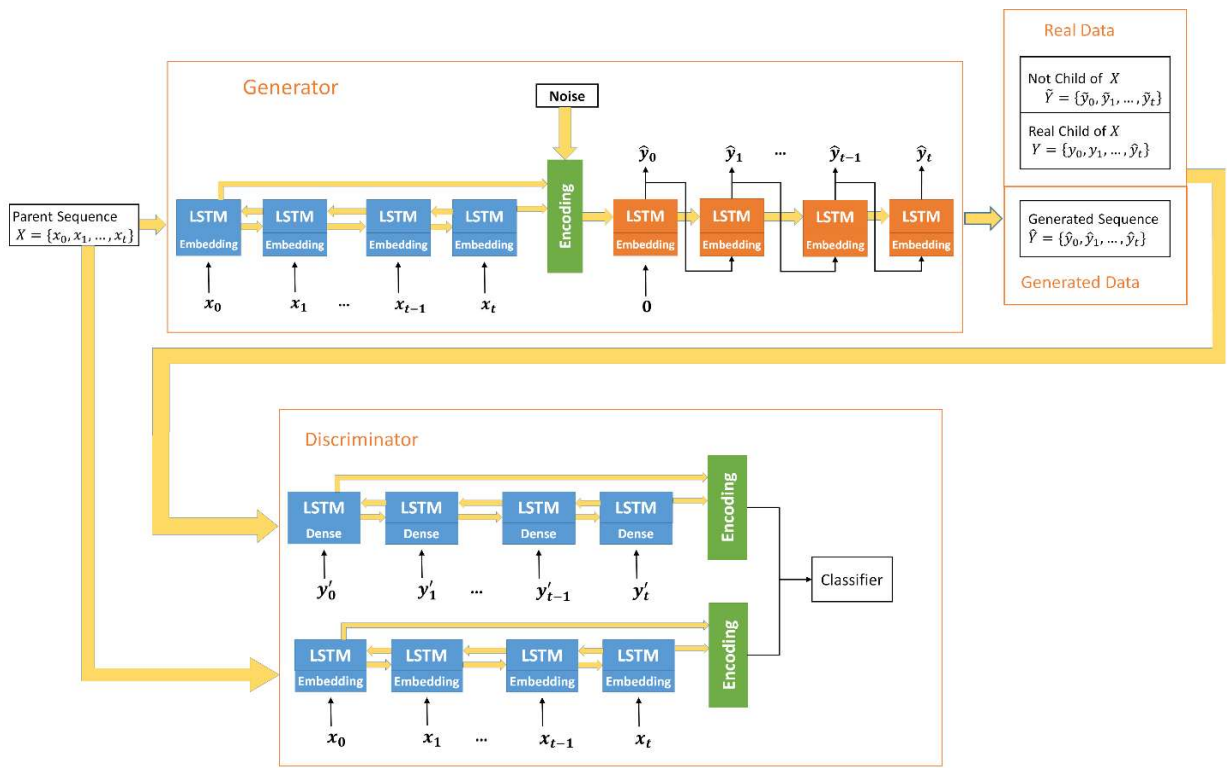

**Figure 2: The MutaGAN framework's architecture.** The generator of the MutaGAN is a seq2seq translation deep neural network using LSTMs and embedding layers. The encoding layer uses a bidirectional LSTM. The output of the encoder is combined with a vector of random noise from a normal distribution $N(0,1)$. The output of the decoder LSTM feeds into a softmax dense layer. An argmax function is then applied to select a single amino acid at each position, rather than a probability distribution. The discriminator uses an encoder with a slightly different structure from the encoder in the generator, but it uses the same weights. This is because an argmax function is not differentiable. Therefore, the first layer of the encoder in the discriminator is a linear dense layer with the output size the same as the term embedding layer in the generator. This allows it to take as input, the output of the dense layer of the decoder in the generator. The weights of this dense layer are the same as those of the embedding layer, meaning it produces a linear combination of the embeddings from the embedding layer of encoder. The discriminator takes in two sequences, and determines whether the input sequences are a real parent-child pair or if it is not. The sequences that are not real parent-child pairs are a parent and generated sequences, and two real sequences that are not parent-child pairs.

## 4 Dataset

For this work, the influenza virus was chosen as an ideal test case for this deep learning framework because it is a significant human pathogen that changes rapidly, with new strains emerging annually, and global surveillance efforts have generated large amounts of publicly available genomic data [46]. The surface proteins HA and neuraminidase (NA) of influenza virus enable virus entry into cells and are the primary immune epitopes that elicit antibodies, making them of particular interest for vaccine development [46].

## 4.1 Database Curation

Influenza virus HA sequences were downloaded from the National Center for Biotechnology Information's (NCBI) Influenza Virus Resource (IVR) [47]. Utilizing Bash text parsing methods including *awk* and *sed*, the dataset was curated to gene sequences from the influenza A type and H3N2 subtype that were obtained from human hosts between 1968 and 2017. Duplicate records were removed using the 'isolate_name' and 'isolation_date' metadata attributes as a unique identifier. When a duplicate identifier was encountered, the first record within the IVR database was kept and the remaining records were discarded. Additionally, only isolates that had at least eight complete gene sequences (i.e. having at least the eight primary protein gene sequences—PB2, PB1, PA, HA, NP, NA, M1, NS1—out of the 12 proteins produced by the influenza virus) present in the dataset were kept. Of note, during curation, 22 isolates from swine or avian hosts remained in this dataset (Supplemental Table 2). When completed, the curated sequence dataset contained 6,840 unique records of H3N2 influenza virus unique sequences.

## 4.2 Phylogenetic Tree Generation

For input into the seq2seq GAN framework, phylogenetic reconstruction was performed using the nucleic acid sequences of the 6,840 HA sequences. The initial approximate maximum likelihood (ML) tree was made using Fasttree [48] with the Jukes-Cantor model and an estimate of 20 per-site rate categories against a multiple sequence alignment of all DNA sequences using MAFFT [49]. This tree was then refined into an intermediate ML tree using RAxML [50] and the rapid hill-climbing mode, a generalized time-reversible (GTR) model, and an estimate of 25 per-site rate categories with the resulting tree evaluated under the GAMMA model of rate heterogeneity model. The final ML tree was made by optimizing the model and branch lengths of the intermediate ML tree using RAxML with a GTR model, GAMMA model of rate heterogeneity, and ML estimate of the alpha-parameter. The final tree, as shown in Figure 3, was visualized using FigTree [51] with some custom post-processing.

## 4.3 Dataset Creation

After rooting the final ML tree to isolate 'influenza A virus A/Hong Kong/1/68(H3N2)', Marginal ancestral sequence reconstruction was performed with RAxML using the General Time Reversible model of nucleotide substitution under the Gamma model of rate heterogeneity. Parent-child relationships were generated using the Bio.Phylo package in BioPython [52] and were limited to single steps between phylogenetic tree levels such that each parent had exactly two children. One parent-child pair was generated for each of the 13,678 edges within the final binary tree. Because phylogenetic tree generation requires removal of duplicate nucleotide sequences prior to evolutionary modeling, there was a concern of providing an information bias of evolution towards the ancestral sequences (i.e. internal nodes) and away from sequences acquired through genomic surveillance (i.e. leaf nodes). To mitigate this bias, leaf nodes that had a nucleotide sequence matching to multiple records within the IVR database were inserted back into the dataset as duplicate parent-child pairs. As an example, if a leaf node's sequence was observed four times in IVR, there would be four identical parent-child pairs inserted into the dataset. Upon completion, the number of parent-child pairs was increased to 17,218 within the formatted dataset. Each nucleotide sequence was translated to amino acids for representation learning of the HA protein by the MutaGAN framework.

The training and test datasets were formed by splitting a list of the compiled unique parent sequences in a random 90/10 split. The result was 1,451 unique parent sequences in the training dataset and 156 unique parents in the test dataset. There were a total of 15,699 parent-child pairs in the training dataset and 1,519 parent-child pairs in the test dataset. A total of 150 sequence pairs (0.96%) were removed from the training dataset and 11 sequence pairs (0.72%) were removed from the test dataset in which the amino acid Levenshtein distance (see Section 5 Generator Evaluation for description) was 10 or greater to prevent parent-child pairs that were excessively unrelated, either from sampling bias or mistaken sequence inclusion. This removal had the effect of isolating an outlier group identified within our phylogenetic tree that appeared as a result of a small number of sequences not being removed during the pre-phylogenetic filtering process. Because the phylogenetic tree was created using the virus gene sequences and synonymous mutations do not lead to amino acid mutations in proteins, the corresponding parent-child protein pairs could be identical. Of the 1,451 unique parents in the training dataset, 103 parents (7.10%) only had child sequences that were identical to the parents, while 567 (39.08%) had only one unique child. For a measure of parent-child diversity, the training set contained 5,048 parent-child pairs where the child's sequence differed from its parent. Matching parent-child pairs were removed from the training dataset. In the test set, all instances of matching parents and children were removed, leaving 433 parent-child pairs with 141 unique parent sequences. The test dataset only contained pairs in which the parent and child sequences were different.

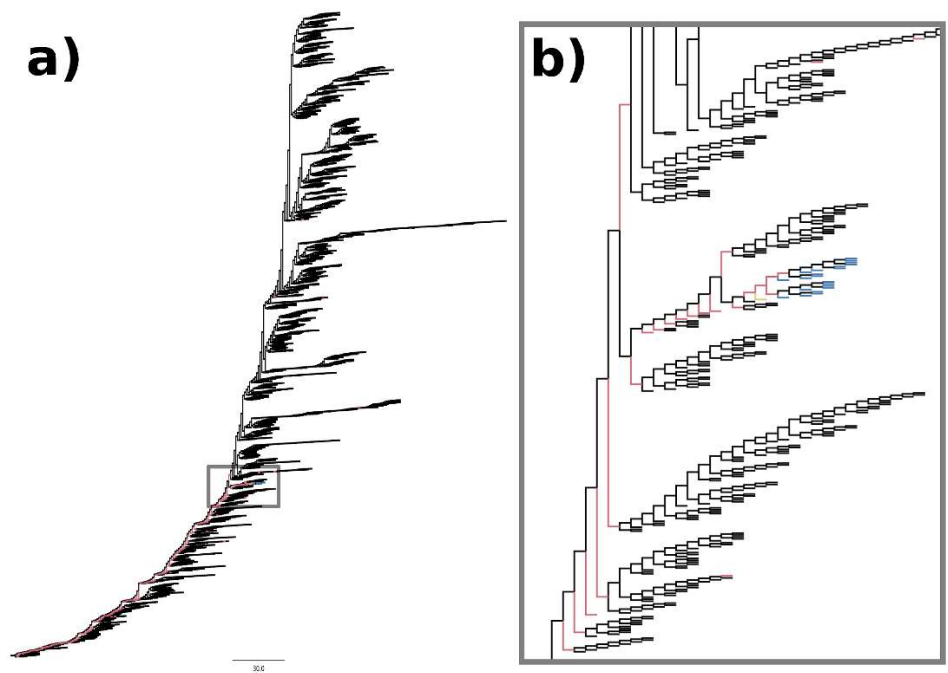

**Figure 3: Topology of RAxML tree used build parent-child pairs.** The topology of the maximum likelihood tree created from 6,840 H3N2 sequences is shown in (a). The ancestral sequences of each internal

node in this tree were used to form the 13,768 parent-child pairs used to train the sequence-to-sequence generator of the GAN framework. An outlying group, containing 22 sequences, was identified as coming from swine or avian hosts, and those sequences are indicated in blue. One of these 22 sequences was from the group of 155 parent-child pairs with a Levenshtein distance greater than 10 and is indicated in yellow. The region surrounding that outlying group (gray box) is expanded in the inset (b), where it can be seen that the majority of the parent-child pairs removed for high Levenshtein distance in salmon come directly off the backbone of the phylogenetic tree leading to the outlying group in blue. This trend continues back to the root of the tree.

## 5 Generator Evaluation

The most important metric for assessing quality of the generated sequences is whether or not they were able to produce the mutations observed in the data. However, missing an observed mutation does not necessarily mean the generator did not correctly predict possible mutations, only that it did not correctly predict all observed mutations. It is possible that predicted mutations would have likely occurred but just weren't included in our subset.

We identified mutations in our sequences by measuring the Levenshtein distance between parent and child sequences. By using Levenshtein distance, we were able to account for insertions and deletions as well as mutations [53], and by using the diff-match package from Google, we were able to identify where changes were made between two sequences [54]. The diff_cleanupSemantic function in the diff-match-patch package was used to identify where changes were made between parents, real children, and generated children. A list of all child mutations was created for each parent and compared to each parent's generated children. A mutation was counted as correct if change occurred in the generated child that was identical in amino acid and location as observed in the parent's real children. A partially correct mutation was defined as an amino acid change in the generated child that was identical in location to any mutation in a real child of that parent but differed in amino acid type. False mutations were defined as mutations that were predicted but non-existent in the real children, and missed mutations were defined as mutations observed in at least one of the real children of the parent but not in the generated child.

Frequency of amino acid mutations was calculated to evaluate the similarity of the mutation profiles between MutaGAN and the ground truth data (Equation (*1*). For each mutation within a given set of mutations, $A_P$ is the amino acid of the parent, $A_C$ is the amino acid of the child, and $x_{A_P A_C}$ is the count of all mutations observed from one amino acid to another. Using this calculation of mutation frequency, the distance between two different sets of mutations is calculated by taking the average of the absolute difference for each amino acid mutation pair between the two data sets:

$$f_{A_P A_C} = \frac{x_{A_P A_C}}{\sum x_{A_P A_C}}. \tag{1}$$

We use two different true positive rates to evaluate the generated sequences. The first is the standard true positive rate, where we compare the amino acids at known mutation locations in the generated and child sequences. We refer to this as the standard true positive rate. The true positive rate is a useful metric in determining if all of our generated sequences contain mutations that match the ground truth data. However, it can be overly harsh in terms of assessing our generated sequences. This is because it penalizes sequences with multiple known mutations where only a portion is present in each generated sequence. If those known mutations are spread across multiple generated sequences, MutaGAN would be penalized for not creating linear combinations of mutations. These linear combinations would be counter to the goal of MutaGAN, which is to simulate the potential evolutionary paths of a protein sequence in a stochastic fashion. For example, a parent sequence ACFKLM has two children, ACFHLM and ACFHIM. If MutaGAN generates 100 sequences, 50 of them are exact matches to child 1, and 50 of them are exact matches to child 2, the true positive rate would be 50.0% because only half of known mutations appear across all generated sequences. If all generated sequences were ACFHIM, the true positive rate would be 100.0% because all known mutations appear in all children. However, we want our generated sequences to be more similar to the former scenario than the latter. Therefore, we can create a true positive rate in which the number of times a mutation was made or missed is ignored, paying attention only to whether or not it happened. We refer to this as the sequence true positive rate because it focuses on the presence of mutations across a given parent sequence. The sequence true positive rate is calculated the same way the standard true positive rate is calculated, with one additional step: a unique list of all the mutations is made for sequences generated for each parent, rather than using the mutations made for each generated sequence.

To account for different levels of similarity between any two amino acids when evaluating mutational errors, Sneath's index [55], a percentage representation of the number of dissimilar comparisons of amino acids along 134 categories of activity and structure was incorporated into a calculation for weighted accuracy. For this analysis, we removed the prediction of the ambiguous amino acid designation (X). Additionally, we set the lower limit on the allowable similarity to 0.85 to prevent over rewarding when calculating weighted averages. Only 18 different amino acid pairings with Sneath's Index have similarities greater than or equal to 0.85 and were included, while all other comparisons were set to 0. This means there are only 18 different types of mistakes for which partial credit can be awarded. To calculate the weighted accuracy, each mutation found in the set of generated children was weighted using the thresholded Sneath's index, $S$, and averaged across the entire table of predicted mutations, $A$, where the columns are the predicted amino acid and the rows are the expected amino acid, as calculated in Equation (2).

$$avg_w = \frac{sum([S \otimes (S \geq 0.85)] \otimes A)}{sum(A)}, \qquad (2)$$

where $\otimes$ is an elementwise multiplication of two matrices with the same dimensions.

A common metric for evaluating text generation is the bilingual evaluation understudy (BLEU) metric [56]. The BLEU metric ranges from 0 to 1, and calculates the precision using the modified n-gram precision and accounts for the micro-average precision of the 1, 2, 3, and 4-grams. The details of how this is calculated are given in [56]. We used the BLEU metric as an additional evaluation because it provides an analysis of the generated sequences from a larger perspective than single amino acids.

## 6 Experiment

In this section, we present the experiment we designed to train and test MutaGAN.

### 6.1 Setup

The phylogenetic tree reconstruction took place on a 16 processor 64GB RAM compute node running Ubuntu. RAxML tree optimization and ancestral reconstruction took roughly 14 days to complete. The models were built and training and testing was implemented in Python version 3.6.3 using the libraries Tensorflow-gpu version 1.8.0 [57] and Keras version 2.1.6 [58] on four GeForce GTX 1080 Ti graphical processing units (GPUs). Additionally, metrics were calculated using the functions in the package scikit-learn version 0.20.3 [59], the diff-match-patch package [53], and the NLTK package [60].

### 6.2 Model Training

The maximum number of words included in the embedding layer was 4500, which was selected by rounding up the number of unique 3-mers found in our dataset. Additionally, we selected an embedding size of 250. The encoder portion was a bidirectional LSTM with 128 nodes, resulting in a state vector of 512 being passed to the decoder, which was a unidirectional LSTM. Generator pretraining was done on the training dataset and tested on the test dataset. This was done using the Adam optimizer [61] with a learning rate of 0.01 until the model reached a stable state. It was set to train for 72 epochs, but converged far before that.

GAN training occurred in two stages based on batch size, for a total of 350 epochs. In both stages, we selected the Adam optimization algorithm and the learning rate for the generator was 1e-3 and the discriminator was 3e-5. The learning rate for the discriminator was chosen to avoid mode collapse. The first 200 epochs of the model were trained on a batch size of 32 with the discriminator for five epochs and the generator training for five epochs. The last 150 epochs of the model were trained on a batch size of 45 with the discriminator and the generator training for five epochs each.

Typically, the discriminator is only meant to help the generator create realistic data, but the MutaGAN discriminator has the added goal of making sure the generated sequence is a possible child of the parent. Therefore, we modeled our approach after Reed et al. [31] and created three types of sequence pairs to train the discriminator. The first pair type is real parents and real children, as determined from the phylogenetic model. The second is real parents and generated children. The third is real parents and real non-children. The purpose of the third pair is to ensure the model learns to differentiate between related and unrelated sequences in the context of evolution. Ten thousand training records of the third type were generated by randomly pairing unrelated parent and child sequences with a Levenshtein distance greater than 15.

To optimize performance of the model, our framework deviated from previously published methods in a number of ways. The MutaGAN seq2seq model was pretrained prior to input into the GAN using teacher forcing [62], so the generator's decoder also contained a similar embedding layer with 4500 words and an embedding size of 250. The loss function was the standard sparse categorical cross entropy loss function.

The initial version of the GAN used a binary cross entropy loss function (Equation (3)).

$$L = \frac{1}{N} \sum_{n=1}^{N} -(y_n \log(p_n) + (1 - y_n) \log(1 - p_n)). \tag{3}$$

However, early iterations of our model using this loss function were characterized by mode collapse, where the generator produces an unvarying child sequence given a single parent sequence. To resolve this problem, the loss function was switched from binary cross entropy to Wasserstein loss

$$L = \frac{1}{N} \sum_{n=1}^{N} y_n * \hat{y}_n. \tag{4}$$

where $y_n$ is the ground truth value, either 1 or -1, and $\hat{y}_n$ is the predicted value, and the final layer of the discriminator into a linear activation function [63]. The loss of the generator is the sum of the Wasserstein loss and the sparse categorical cross entropy of generating the child sequence.

In a variation from Reed et al. [31], we used sequences generated by the initial GAN as additional negative examples in training the discriminator of the final model to prevent the model from drifting too far off course as a form of experience replay, similar to an approach used in deep reinforcement learning [64]. The initial GAN created a high proportion of generated children with a Levenshtein distance greater than 300 (Supplemental Figure 1). Using this model to generate 10,000 children from randomly selected parents, with replacement, and removing pairs where the Levenshtein distance was less than 15, we were left with 8,550 parent-child pairs. These sequences were used for experience replay. The distribution of the Levenshtein distances of the sequences for both the fake parent-child pairs and the experience-replay pairs is shown in Supplemental Figure 1 in a stacked histogram, with the real parent and real non-child sequences in blue, and the real parent and generated sequences from the failed model in orange.

## 7 Results

After the model was trained, we generated 100 child sequences for each of the 141 unique parent sequences in the test set. We then discarded 10 sequences with greater than 60 amino acid differences from their parent sequences, a non-biological artifact that we are treating as noise. A change of this magnitude corresponds to 10% of the overall protein structure and is highly improbable to have occurred by chance within a single evolutionary step on the timescale with which the phylogenetic model was created. Given this context, we felt comfortable removing them as this is how these sequences would likely be treated in real life applications of MutaGAN. The median Levenshtein distance between parent and observed child amino acid sequences within the test set was 1.00 ($\sigma = 1.06$). The median Levenshtein distance between the generated sequence and the parent sequence is 2.00 ($\sigma = 4.56$) (Supplemental Figure 2). The generated sequences are very close, but not identical to the parent sequences, indicating that the model is augmenting its input to account for the learned model of protein evolution.

Because amino acids range in biochemical and physical similarities, it is important to look closer at the actual mutations that are made or missed, especially because many of the "mistakes" correspond to common biological mutations between functionally similar amino acids. Mutation profiles by amino acid are provided in Figure 4. A side-by-side comparison of the mutational profiles is made across the training, test, and MutaGAN-generated amino acid sequences with respect to the parent input sequences. MutaGAN's amino acid mutational profile is strikingly similar to that of both the training and test datasets, indicating that the model has learned a measure of biological significance in the bio-physical and chemical properties of amino acids. To assess if MutaGAN's generated amino acid mutation profile more closely resembles the training set over the test set or vice versa, the average difference in amino acid mutation frequency was calculated for the "Generated vs. Training" and "Generated vs. Test" delta mutation profiles (Figure 4B). This measure of distance was calculated to be 1.42E-3 and 1.38E-3, respectively. A comparison between the two distances shows that MutaGAN's mutation profile is closer to the test set by only 4 mutations in 100,000 and therefore, it is concluded that the generated amino acid mutation profile is not significantly closer to one ground truth set over the other. Importantly, however, the MutaGAN generated mutation profile shows changes in mutation frequency for specific amino acids that more closely resembles the test set when compared to the training set (Figure 4A). In particular, it is observed that there are higher proportions of Threonine (T)→Lysine (K), Threonine (T)→Isoleucine (I), Arginine (R)→Lysine (K), and Glycine (G) →Aspartic Acid (D) mutations and lower proportions of Alanine (A) →Valine (V), Glycine (G) →Arginine (R), and Alanine (A) →Threonine (T) within the ground truth test data as compared to the ground truth training data. For these same amino acid, MutaGAN's mutational profile shows the same trends.

The most prominent amino acid mutations that were made by MutaGAN that were not seen frequently in either the training or test data are Glutamine (Q)→Arginine (R), Asparagine (N)→Aspartate (D), Asparagine (N)→Serine (S), and Histidine (H)→Asparagine (N) (Figure 4B). Interestingly, Arginine is the second most favorable amino acid mutation from Glutamine behind Glutamate [65]. Aspartate and Serine are the most favorable amino acid mutations from Asparagine alongside Histidine and Asparagine is the second most favorable mutation from Histidine behind Tyrosine. Another frequently incorrect MutaGAN mutation of note is Serine (S)→Proline (P). Serine, when present on a protein's surface, often forms hydrogen bonds with the protein's backbone and effectively mimics Proline [65]. In accordance, the four locations that MutaGAN incorrectly mutated the hemagglutinin protein from a Serine to a Proline were at amino acid positions 143, 198, 199, and 227 within the HA1 chain, all of which are located on protein's surface.

As an artifact of the phylogenetic analysis, a small but noticeable portion of child sequences within both the training and test datasets contained the ambiguous amino acid symbol 'X' at some location within its sequence. The appearance of 'X' in a child sequence created the appearance that a parent amino acid could mutate to ambiguity. However, MutaGAN never mutated a parent amino acid to ambiguity (Figure 4A). This is likely due to the fact that of all the amino acids in the training dataset, only 3.35E-3 percent were 'X', meaning there were too few examples of 'X' for it to learn it.

The overall mutation location profile of historical H3N2 influenza virus hemagglutinin proteins was well reproduced by MutaGAN (Figure 5). The most highly variable regions identified in the training and test datasets (HA1 amino acid indices 120-160 and 185-228) were also the most mutated regions by MutaGAN. Regions of lesser, but still significant, variability were also identified by MutaGAN in accordance with the historical H3N2 data observed in the training and test sets of Figure 5 such as HA1 residue regions (i.e. amino acid indices) of 45-59, 259-262, and 273-278. Regions of historical conservation were accurately preserved by MutaGAN, most notably the HA1 residue region of 11-24 and the HA2 residue region 1-16. Of the top ten most frequently mutating positions in the training dataset, MutaGAN's only had one within its own top ten (position 121). Between the test set and MutaGAN-generated set, there were no overlaps in the top ten most frequently mutated positions. However, for many of the most frequently mutating amino acid locations within the training and test sets, the location was one or two positions away from a commonly mutated position in the MutaGAN generated sequences. For instance, HA1 residues 142, 160, and 193 were in the top ten most frequently mutated positions in the training and test sets. HA1 residues 145, 159, and 192 were in the top ten most frequently mutated positions by MutaGAN. This phenomenon is worth noting because of the closeness, but the biological significance is not readily apparent without a deeper analysis of hemagglutinin protein structure. In looking at the structure of the hemagglutinin protein more closely (Figure 5B), it is clear that the concentration of the most frequently mutated position for both the training and test data sets occurs on the outside of the protein structure, principally on the outer surface of the HA1 domain towards the host-recognition regions. It is well understood that the frequent mutation of amino acids at these locations increases the influenza virus's likelihood of evading host antibodies during infection. The majority of the MutaGAN-generated mutations also occur in these same regions but across a notably larger number of residues on the protein surface. This finding alludes to the ability of this framework to illuminate localized function across varying regions of the overall protein structure but further simulations must be performed to investigate the functional effects of the MutaGAN-generated mutations.

While recognizing that the phylogenetic tree does not capture the entire breadth of mutations that occurred during the entire evolution of the influenza virus, MutaGAN's performance was evaluated with respect to this tree as our closest proxy of its ability to mimic the virus's evolutionary landscape. The number of observed mutations reproduced for each parent is visualized in Figure 6 and shows that most generated children contained at least one observed mutation, while a smaller number contained more. 93 of the 141 parent sequences (66.0%) had at least one observed mutation augmented onto it within its generated child sequences. Of the parent sequences that MutaGAN did not correctly identify a mutation as observed in the ground truth, there were seven (4.96%) sequences for which MutaGAN produced a mutation in the correct location but with the incorrect amino acid.

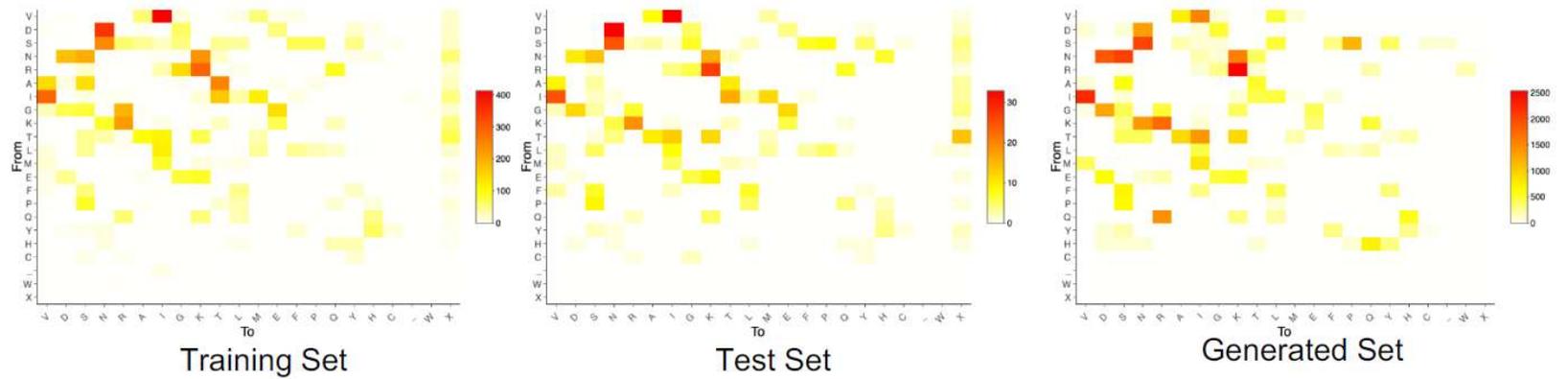
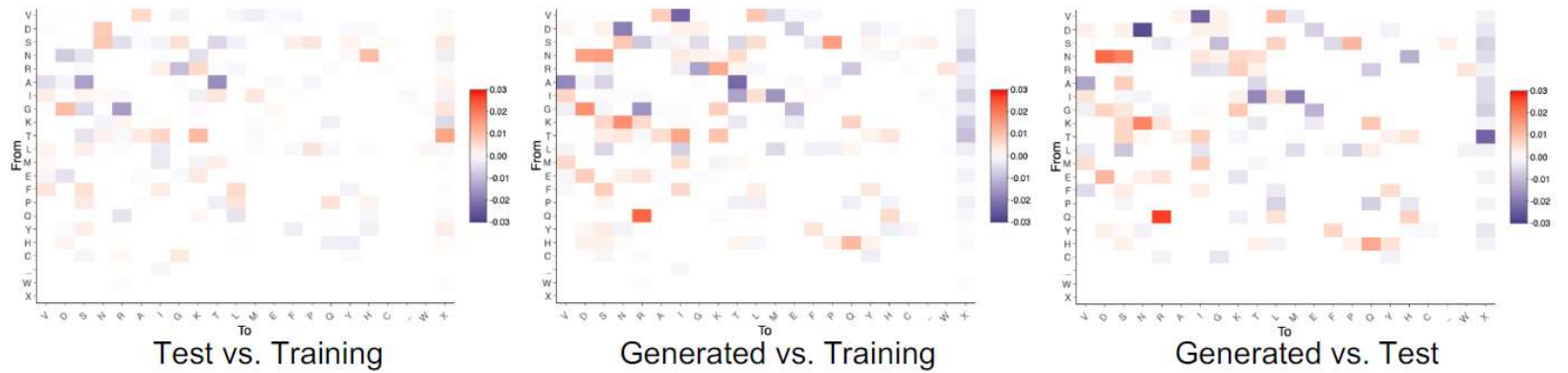

**Figure 4: Amino acid mutation profiles with respect to amino acid types.** For the training, test, and generated child sequences, total counts for each amino acid mutation from parent to child are displayed in (a). Amino acid ordering was determined using R's hclust function on the training data and kept consistent throughout both (a) and (b). Differences in amino acid mutation frequency between the training, test, and generated datasets was calculated and visualized in (b) using Equation 1.

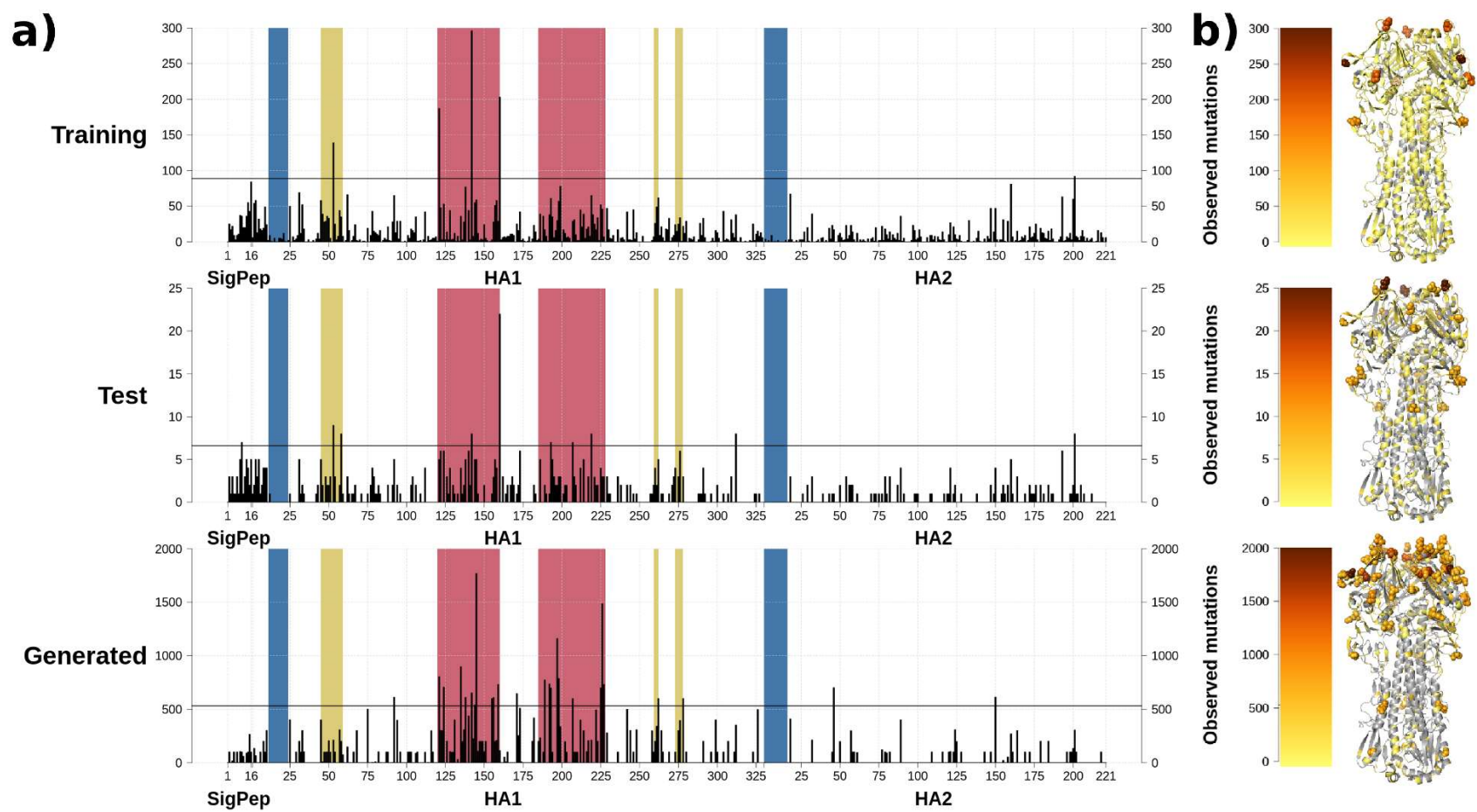

**Figure 5: Amino acid mutation profiles with respect to HA protein locations.** For the training, test, and generated child sequences, total counts of mutations observed across the entire length of the HA protein segment are displayed in (a), indicating the signal peptide, HA1 (head), and HA2 (stalk) regions of the full hemagglutinin protein. The most highly variable regions are highlighted in salmon. Regions of lesser, but still significant variability, are highlighted in yellow. Particularly conserved regions are highlighted in blue. In (b), a diagram of the H3 hemagglutinin structure (PDB: 4GMS) is colored by this mutation frequency, with the positions with the fewest mutations in yellow to the positions with the most mutations in brown. Positions with zero observed mutations across each dataset are colored gray. Residues are displayed as spheres for positions with mutation frequencies above 30% of the maximum position for each of the three datasets. These 30% threshold lines are also plotted in (a).

As a measure of the standard true positive rate, each generated child's sequence contained 19.7% of observed mutations between its parent and real children (the standard true positive rate comparing the generated and real child sequences). Using Sneath's index [55] to get a deeper assessment of the closeness of MutaGAN's predictions, we find that it has a weighted true positive rate of 57.7%, calculated using equation (*2*, to compare the generated and real child sequences. This large increase from an unweighted standard true positive rate of 19.7% to 57.7% indicates that a majority of the mutations found within the real child sequences are similar in biochemical and physical properties to the amino acids MutaGAN used in those location. As a measure of false negatives, 75.7% of the mutations that MutaGAN predicted were not observed in the test set. Weighting this using the Sneath's index, this drops to 39.3%, indicating most of the mutations MutaGAN made are similar in biochemical and physical properties to amino acids in those locations in the real child sequences. Acknowledging again that some of these mutations could have occurred in reality and were not captured by genomic surveillance efforts, the high false positive and false negative rates indicate that MutaGAN could benefit from an improved phylogenetic model and additional training data to tune the model's predictions.

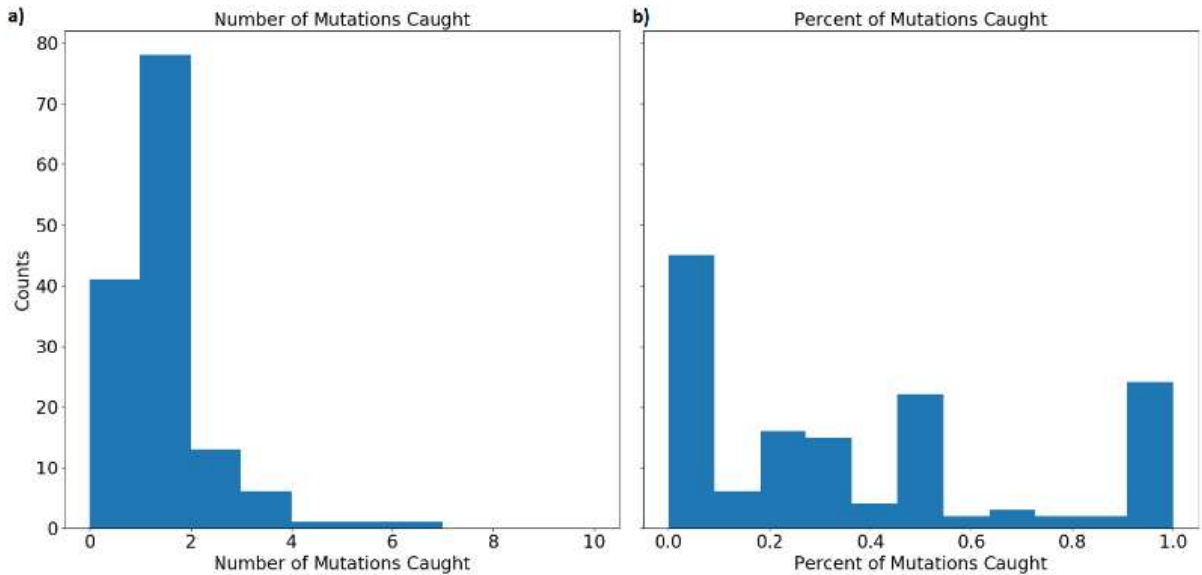

**Figure 6: Histograms showing the distribution of correct mutations** as both total counts (a) and percentage of total recorded mutations (b).

When we use the sequence true positive rate, we find the unweighted sequence true positive rate is 21.0% with an 84.8% false positive rate, and the weighted sequence true positive rate is 64.1% and weighted false positive rate is 47.1%, using the same weighting scheme as above. The BLEU score for our model is 97.46%, which indicates a high level of precision in sequence generation.

For 81 parent sequences (57.4%), our model generated the same child sequence in each of the 100 iterations, regardless of the noise, using the distribution *N(0,1)* (Supplemental Figure 3). While this behavior might appear to be mode collapse, in the ground truth test data, the phylogenetic tree had only one child for 79 (56.0%) of the parent sequences as a direct result of using the amino acid rather than nucleotide sequences and masking synonymous mutations. Of these 81 parent sequences whose generated children exhibited mode collapse-like behavior, 40 (49.4%) also had only one child in the ground truth phylogenetic tree data. Since these proportions of parents with only one child are similar in both the generated results and the test dataset, and notably larger than the 39.1% in the training dataset, and there is a large overlap in sequences with only one child in the generated results and test dataset, it appears that the mode collapse is only partially responsible. Future work, such as operating directly on nucleotide sequences, could help reduce the impact of mode collapse on the sequence generation.

# 8 Conclusion

To our knowledge, the MutaGAN framework is the first method to utilize a GAN to accurately reproduce and optimize full length proteins above 300 amino acids in length with no structural information provided to the model beyond amino acid sequence. The accurate reproduction of the mutation profiles of the HA protein with specificity of both the types of amino acids changed (Figure 4) and locations most likely for persistence (Figure 5) demonstrates the potential of this method to be used as a tool for forecasting the genetic drift or shift that occurs during the outbreak of the influenza virus. Our findings indicate that the sequence augmentation strategies deployed by MutaGAN optimize its input towards the most successful patterns observed during the evolutionary history of a protein. Because this framework is agnostic to the type of phylogenetic tree and protein type used to generate parent-child pairs, the extension of these methods to new proteins and organisms (e.g. the NA protein for influenza and the dengue virus) is ripe for exploration.

The ability of MutaGAN to learn and optimize mutations for persistence within a population lends itself well to protein engineering applications. As demonstrated in Figures 4 and 5, the unique nuances of change within a protein population are capable of being captured without any additional expert knowledge being provided to the model beyond a list of parent-child pairs for training tailored to a specific protein. The observation that MutaGAN inserts mutations that are biologically relevant, even when not observed in the ground truth data, poses the question of whether these mutations produce energetically favorable protein conformations with increased fitness within the evolutionary landscape. Future work could pair computational protein modeling with this framework for a deeper analysis of the MutaGAN-generated sequences for improved forecasting of population-level mutation propagation. With direct ties to the public health domain, by measuring the conformational protein favorability of MutaGAN-generated sequences and analyzing their similarity to currently circulating pathogenic sequences, public health officials could assess the threat of potential mutations against vaccine evasion and improve the design of future treatments or vaccines.

Although the mutation profiles are well-reproduced with respect to amino acid and location, MutaGAN's performance, when evaluated at the single nucleotide resolution, has significant room for improvement. The 19.7% and 21.0% capture of standard and sequence true positive rates and 75.7% and 84.8% prediction of false positives highlights shortcomings of the current trained model. It is hypothesized that the model could be improved solely by improving the dataset. There is a high likelihood that the inclusion of swine and avian influenza sequences into the phylogenetic model inhibited the MutaGAN's ability to fit itself on patterns of HA protein evolution specific to human infection. In addition to a more comprehensive curation of outliers, a larger population of HA protein sequences could be utilized to provide additional diversity to the model. The small number of database records (6,840) used to generate the phylogenetic model was unlikely to capture the full breadth and depth of the true evolutionary landscape of the human H3N2 influenza virus HA protein. By leveraging a larger database of influenza virus surveillance, such as GISAID's EpiFlu [66], a more complete evolutionary model could be generated and provided to MutaGAN for training and validation. In addition to the diversity of sequences provided to the phylogenetic tree, constructing this phylogenetic model using time-based Bayesian tree estimation methods could improve the ability of MutaGAN to learn time-related aspects of evolution as well as enable a deeper characterization of the model's ability to forecast into the future. This approach could also provide us the opportunity to compare our model predictions to those of the experts' predictions in a given year's influenza vaccine. Future studies will also explore non-deterministic methods of ancestral sequence reconstruction utilizing the simulated nucleotide probabilities per position rather than strictly including the most probable sequence of the internal node for inclusion in the parent-child pair.

There are also plans to further improve upon the MutaGAN framework's architecture. There are a number of recent advances in NLP and sequence generation that can be leveraged to further improve this algorithm. These advancements include models like attention [67] [68] [69], BERT [70], and reinforcement learning [32] [34] [35] [71]. These models have shown significant improvement over models relying on LSTMs alone for NLP tasks. When paired with the larger dataset provided by a larger influenza database and non-deterministic methods for ancestral sequence reconstruction, we believe the fidelity of sequence reconstruction and optimization can be improved. Because operating directly on nucleotide data increases the length of the sequence data from amino acids by a factor of three and further inhibits the RRN identification of long-range structural relationships, it was avoided in this study. However, with the implementation of more robust encoder-decoder architecture, future research could evaluate the feasibility of MutaGAN to operate directly on nucleotide sequences. Doing so would align MutaGAN with the industry standard in phylogenetic analysis and potentially enable improved learning of evolutionary landscapes through the added information of synonymous mutations observed within protein lineages.

Taken together, we have developed a first-of-its-kind deep learning framework to predict genetic evolution in dynamic biological populations. As a result, we see the potential for this research to play a significant role in public health, particularly in disease mitigation and prevention. With the improvements outlined above, if MutaGAN was implemented to simulate how currently circulating pathogens could evolve over time, targeted measures of quarantine and treatment could be more effectively deployed. MutaGAN's ability to produce full length protein sequences while simultaneously learning the nuances of evolution lends itself well to the extension to other protein types, creating potential for impact within the domain of multiple diseases.

**Acknowledgements**: We are grateful to Jason Fayer, and Ben Baugher for their support, comments, corrections, and feedback.

**Funding**: This research was funded in part by the JHUAPL Janney Program and a contract from the National Institute of Allergy and Infectious Diseases (NIAID) Centers of Excellence in Influenza Research and Surveillance (HHSN272201400007C).

**Conflicts of Interest**: The authors declare no conflicts of interest.

# Appendix A. Supplemental Information

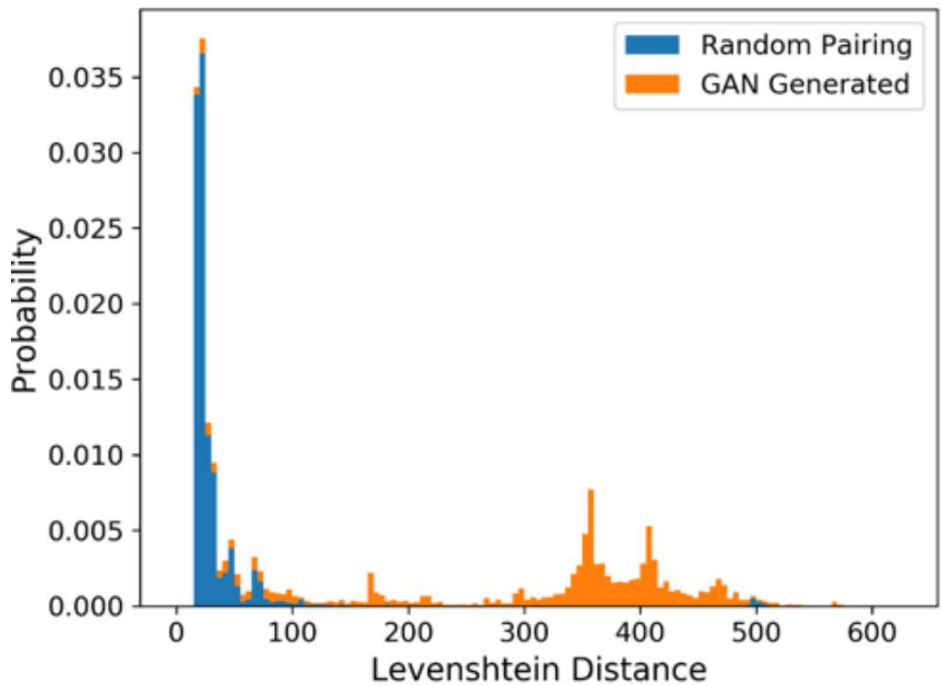

**Supplemental Figure 1: Distribution of Levenshtein distance between parent and non-related child, and a parent and generated child from a previous model.**

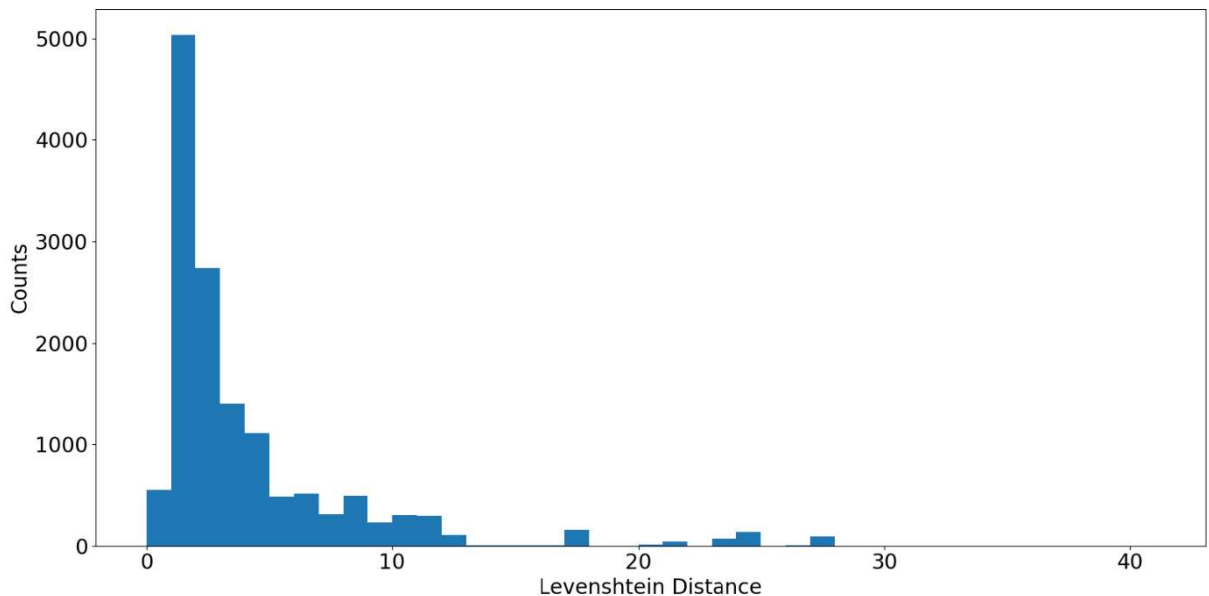

**Supplemental Figure 2: Histogram of Levenshtein distances between parent and generated sequence.**

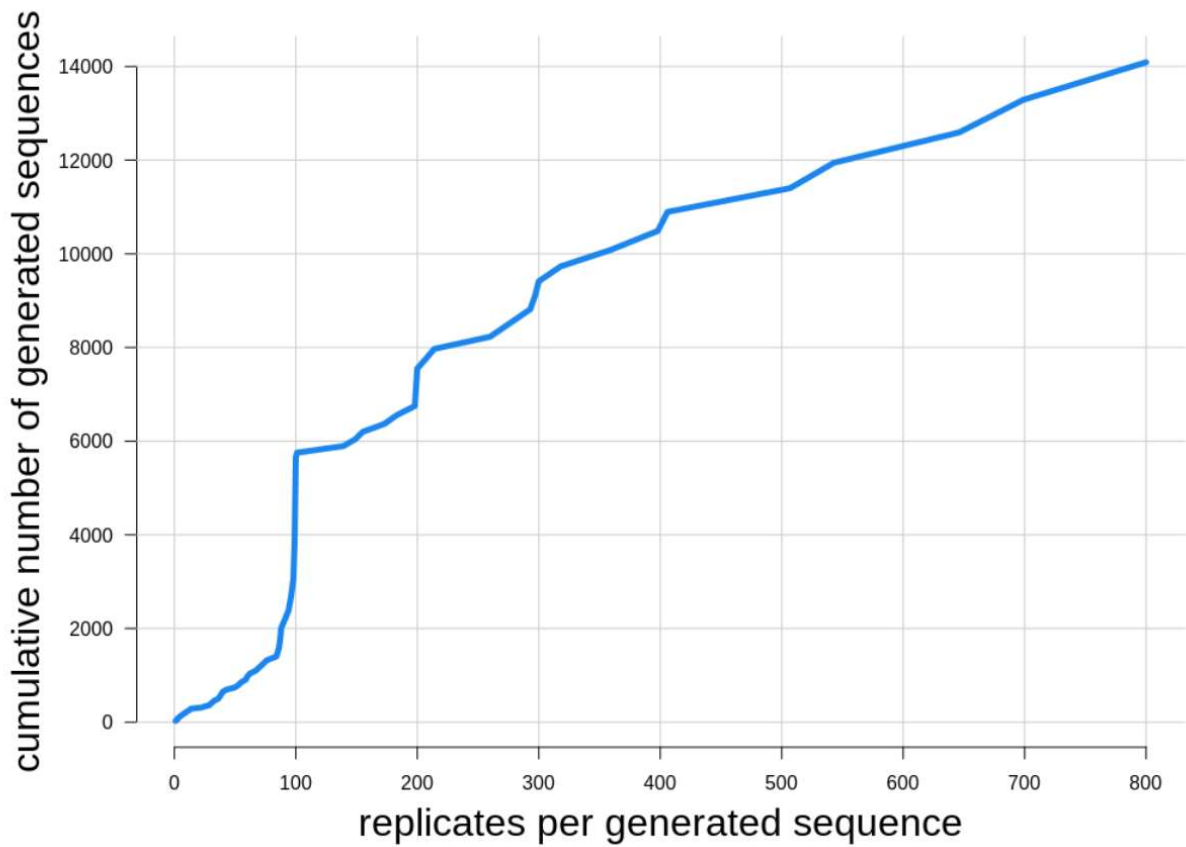

**Supplemental Figure 2: Plot of the cumulative number of generated sequences as a function of replicates per sequence.**

```python
#Parent Sequence Encoder
par_inputs = Input(shape=(None,))
enc_embedding = Embedding(4500, 250)
par_inputs_emb=enc_embedding(par_inputs)
enc_BiLSTM = Bidirectional(LSTM(128,return_state=True))
enc_outputs, fwd_hPar, fwd_cPar, bwd_hPar, bwd_cPar = enc_BiLSTM(par_inputs_emb)
state_hPar = Concatenate()([fwd_hPar, bwd_hPar])
state_cPar= Concatenate()([fwd_cPar, bwd_cPar])
encoder_statePar = [state_hPar, state_cPar]
ParentEncoder_model = Model([par_inputs], encoder_statePar)

#Generated Sequence Encoder
gen_inputs = Input(shape = (None,4500))
enc_inputsGen = Dense(250, activation = 'linear', use_bias = False)(gen_inputs)
enc_outputsGen, fwd_hGen, fwd_cGen, bwd_hGen, bwd_cGen = enc_BiLSTM(enc_inputsGen)
state_hGen = Concatenate()([fwd_hGen, bwd_hGen])
state_cGen = Concatenate()([fwd_cGen, bwd_cGen])
encoder_stateGen = [state_hGen, state_cGen]
GenEncoder_model = Model([gen_inputs], encoder_stateGen)

#Load the weights for the pretrained autoencoder
ParEncoder_model.load_weights('Influenza_biLSTM_encoder_model_128_4500_weights.h5')
GenEncoder_model.layers[1].set_weights(enc_embedding.get_weights())

xPar = ParEncoder_model([par_inputs])
xGen = GenEncoder_model ([gen_inputs])
xConcat = Concatenate()(xPar+xGen)
x = Dropout(0.2)(XConcat)
x = BatchNormalization()(x)
x = Dense(128)(x)
x = LeakyReLU(0.1)(x)
x = Dropout(0.2)(x)
x = BatchNormalization()(x)
x = Dense(64)(x)
x = LeakyReLU(0.1)(x)
x = Dropout(0.2)(x)
x = BatchNormalization()(x)
output_class = Dense(1, activation = 'linear')(x)
Discriminator = Model([par_inputs, gen_inputs], output = [output_class])
```

**Supplemental Table 1: Python code for function that built the discriminator portion of the GAN.**

| Accessions |
|---|
| 1. LC106066 |
| 2. LC106067 |
| 3. LC106068 |
| 4. LC106069 |
| 5. LC106070 |
| 6. LC106071 |
| 7. LC106072 |
| 8. LC106073 |
| 9. LC106074 |
| 10. LC335983 |
| 11. MG912581 |
| 12. MG964344 |
| 13. MG964360 |
| 14. MG964368 |
| 15. MG964384 |
| 16. MG964408 |
| 17. MG964416 |
| 18. MG964440 |
| 19. MG964448 |
| 20. MG964456 |
| 21. MG964488 |
| 22. MG964512 |

**Supplemental Table 2: List of the NCBI accessions of the outlying group that were mistakenly included in the dataset.**